\documentclass[12pt]{article}

\usepackage{geometry}
\usepackage{graphicx}
\usepackage[utf8]{inputenc}
\usepackage{amsmath,amscd}    % need for subequations
\usepackage{graphicx}   % need for figures
\usepackage{verbatim}   % useful for program listings
\usepackage{color}      % use if color is used in text
\usepackage{subfigure}  % use for side-by-side figures
\usepackage{hyperref}   % use for hypertext links, including those to external documents and URLs
\usepackage{bbm}
\usepackage{amsfonts}
\usepackage{mathtools}
\usepackage{amssymb}
\usepackage{tensor}
\begin{document}

%\maketitle
\author{Niels G. Gresnigt}

\title{The Standard Model particle content with complete gauge symmetries from the minimal ideals of two Clifford algebras}
\maketitle

\abstract{

Building upon previous works, it is shown that two minimal left ideals of the complex Clifford algebra $\mathbb{C}\ell(6)$ and two minimal right ideals of $\mathbb{C}\ell(4)$ transform as one generation of leptons and quarks under the gauge symmetry $SU(3)_C\times U(1)_{EM}$ and $SU(2)_L\times U(1)_Y$ respectively. The $SU(2)_L$ weak symmetries are naturally chiral. Combining the $\mathbb{C}\ell(6)$ and $\mathbb{C}\ell(4)$ ideals, all the gauge symmetries of the Standard Model, together with its lepton and quark content for a single generation are represented, with the dimensions of the minimal ideals dictating the number of distinct physical states. The combined ideals can be written as minimal left ideals of $\mathbb{C}\ell(6)\otimes\mathbb{C}\ell(4)\cong \mathbb{C}\ell(10)$ in a way that preserves individually the $\mathbb{C}\ell(6)$ structure and $\mathbb{C}\ell(4)$ structure of physical states. This resulting model includes many of the attractive features of the Georgi and Glashow $SU(5)$ grand unified theory without introducing proton decay or other unobserved processes. Such processes are naturally excluded because they do not individually preserve the $\mathbb{C}\ell(6)$ and $\mathbb{C}\ell(4)$ minimal ideals.

}

\section{Introduction}

The Standard Model (SM) is currently the best model of particle physics. It provides a hugely predictive mathematical description of the most basic constituents of matter observed in nature, and their interactions via three of the four fundamental forces of nature. However, despite being able to use the theory to make remarkably accurate predictions, the theoretical origin for the mathematical structure of the model remains unexplained. Specifically, why is the internal symmetry group of the SM what it is, when infinitely many other possibilities exist, and why do only some of the representations of the gauge groups correspond to physical states? Furthermore, why does the electroweak force depend on the chirality of physical states? Among others, these remain important open questions.

Grand Unified Theories (GUTs) merge the gauge groups of the SM into a single semi-simple Lie group. However, such GUTs, including the famous Georgi and  Glashow's $SU(5)$ model and Georgi's  $\textrm{Spin}(10)$ model \cite{georgi1974unity}, invariably predict additional gauge bosons, interactions, and proton decay, none of which have thus far been observed.

In 2016, the basis states of the minimal left ideals of the Clifford algebra $\mathbb{C}\ell(6)$ were shown to transform precisely as a single generation of leptons and quarks under the electrocolor group $SU(3)_C\otimes U(1)_{EM}$\cite{furey2016standard}. The Clifford algebra $\mathbb{C}\ell(6)$ there is generated from the left adjoint actions of the complex octonions $\mathbb{C}\otimes\mathbb{O}$ on itself\footnote{The application of octonions to quark symmetries goes back to the 1970 \cite{gunaydin1973quark}. More recently division algebras have received some renewed interest.  The results of \cite{furey2016standard} were extended to three generations recently by going beyond the octonions, and considering the next Cayley-dickson algebra of sedenions \cite{gillard2019three}. Dixon \cite{dixon2013division} showed that the composition algebra $\mathbb{R}\otimes\mathbb{C}\otimes\mathbb{H}\otimes\mathbb{O}$ plays a key role in the architecture of the SM. A gravitational theory based on a $\mathbb{R}\otimes\mathbb{C}\otimes\mathbb{H}\otimes\mathbb{O}$ metric has been constructed in \cite{perelman2019r}.}. A Witt decomposition of $\mathbb{C}\ell(6)$ splits the algebra into a basis of nilpotent ladder operators, and the unitary symmetries that preserve this Witt decomposition are $SU(3)$ and  $U(1)$. These symmetries act on the minimal left ideals of the algebra, with each basis state of the ideal transforming like a lepton or quark under the unbroken gauge symmetry $SU(3)_C\times U(1)_{EM}$.

Similarly, considering the left and right adjoint actions of the complex quaternions $\mathbb{C}\otimes\mathbb{H}$ each generate a distinct copy of $\mathbb{C}\ell(2)$, and allows one to describe the spin and chirality of leptons and quarks respectively. Combined, the left and right actions give a representation of the Dirac algebra $\mathbb{C}\ell(4)$ \cite{furey2016standard,furey20183}.

In \cite{furey20183} it is shown that the individual $\mathbb{C}\ell(6)$ and $\mathbb{C}\ell(4)$ results can be combined, and that it is possible to represent one generation of chiral fermions with the SM gauge symmetries in terms of $\mathbb{C}\ell(10)$ minimal ideals. However, this particular representation of a generation of fermions as a minimal left ideal of $\mathbb{C}\ell(10)$ does not preserve the individual $\mathbb{C}\ell(6)$ and $\mathbb{C}\ell(4)$ ideal structures associated with specific leptons and quarks. This is the direct result of how the $SU(2)_L$ generators are defined, which do not account for the fact that the two particles in a weak doublet are elements of different $\mathbb{C}\ell(6)$ minimal left ideals. The $\mathbb{C}\ell(6)$ minimal left ideals are invisible to the $SU(2)_L$ generators meaning that the two particles making up a weak doublet must have the same $\mathbb{C}\ell(6)$ structure, and differ only in their $\mathbb{C}\ell(4)$ structure. 

This paper shows that by suitably redefining the $SU(2)_L$ generators in such a way that they simultaneously induce transformations in the $\mathbb{C}\ell(4)$ minimal right ideals as well as map between different $\mathbb{C}\ell(6)$ minimal left ideals, it is possible to to embed all the physical states into minimal left ideals of $\mathbb{C}\ell(10)$ in a particularly aesthetic way that preserves both the individual $\mathbb{C}\ell(6)$ minimal left ideal and $\mathbb{C}\ell(4)$ minimal right ideal basis states associated with leptons and quarks. The unitary symmetries that preserve a Witt decomposition of this larger algebra generate $SU(5)$. However, not all of the $SU(5)$ transformations individually preserve the $\mathbb{C}\ell(6)$ minimal left ideals and $\mathbb{C}\ell(4)$ minimal right ideals. Although such transformations would not be excluded if one starts with $\mathbb{C}\ell(10)$ as the fundamental background structure, as one does in the $SU(5)$ and Spin$(10)$ GUTs, they are naturally excluded here. It is these transformations that correspond to unobserved processes, including proton decay. In the model presented here, such unphysical transformations are therefore algebraically excluded.

Besides giving a particularly aesthetic embedding of one generations of fermions into minimal ideals of $\mathbb{C}\ell(10)$, our model also provides a clear pathway for extending earlier work \cite{gresnigt2018braids} which established a curious structural similarity between the basis states of the minimal left ideals of $\mathbb{C}\ell(6)$ and certain braids used in a topological model of leptons and quarks \cite{Bilson-Thompson2005}. Incorporating the results of the present paper to extend these earlier results to include the weak interaction will be the focus of a future paper.

%%%%%%%%%%%%%%%%%%%%%%%%%%%%%%%%%%%%%%%%%%%%%%%
\section{Electrocolor symmetries for one generations of fermions from $\mathbb{C}\ell(6)$}
In \cite{furey2016standard} it was shown that a Witt decomposition of the $\mathbb{C}\ell(6)$ decomposes the algebra into minimal left ideals whose basis states transform as a single generation of leptons and quarks under the unbroken electrocolor symmetry $SU(3)_C\times U(1)_{EM}$. Such a Witt basis of $\mathbb{C}\ell(6)$ can be defined as\footnote{following the convention of \cite{furey2016standard}.}
\begin{eqnarray}
\alpha_1&\equiv&\frac{1}{2}(-e_5+ie_4),\qquad \alpha_2\equiv\frac{1}{2}(-e_3+ie_1),\qquad \alpha_3\equiv \frac{1}{2}(-e_6+ie_2),\\
\alpha_1^{\dagger}&\equiv& \frac{1}{2}(e_5+ie_4),\qquad \alpha_2^{\dagger}\equiv\frac{1}{2}(e_3+ie_1),\qquad \alpha_3^{\dagger}\equiv \frac{1}{2}(e_6+ie_2),
\end{eqnarray}
satisfying the anticommutator algebra of fermionic ladder operators
\begin{eqnarray}\label{mtis}
\left\lbrace \alpha_i^{\dagger},\alpha_j^{\dagger} \right\rbrace=\left\lbrace \alpha_i,\alpha_j \right\rbrace=0,\qquad \left\lbrace \alpha_i^{\dagger},\alpha_j \right\rbrace =\delta_{ij}.
\end{eqnarray}
The $\alpha_i$ and $\alpha_i^{\dagger}$ are nilpotents and each set $\left\lbrace \alpha_i \right\rbrace$ and $\left\lbrace \alpha_i^{\dagger}\right\rbrace $ generates a maximal totally isotropic subspace of dimension eight. One can then construct the minimal left ideal $S_0^u\equiv \mathbb{C}\ell(6)\omega\omega^{\dagger}$, where $\omega\omega^{\dagger}=\alpha_1\alpha_2\alpha_3\alpha_3^{\dagger}\alpha_2^{\dagger}\alpha_1^{\dagger}$ is a primitive idempotent. Explicitly:
\begin{eqnarray}\label{Ideal1}
\nonumber S^u\equiv &{}&\\
\nonumber &{}&\;\;\nu \omega\omega^{\dagger}+\\
\nonumber \bar{d}^r\alpha_1^{\dagger}\omega\omega^{\dagger} &+& \bar{d}^g\alpha_2^{\dagger}\omega\omega^{\dagger} + \bar{d}^b\alpha_3^{\dagger}\omega\omega^{\dagger}\\
\nonumber u^r\alpha_3^{\dagger}\alpha_2^{\dagger}\omega\omega^{\dagger} &+& u^g\alpha_1^{\dagger}\alpha_3^{\dagger}\omega\omega^{\dagger} + u^b\alpha_2^{\dagger}\alpha_1^{\dagger}\omega\omega^{\dagger}\\
&+& e^{+}\alpha_3^{\dagger}\alpha_2^{\dagger}\alpha_1^{\dagger}\omega\omega^{\dagger},
\end{eqnarray}
where $\nu$, $\bar{d}^r$ etc. are suggestively labeled complex coefficients denoting the isospin-up elementary fermions. The conjugate system analogously gives a second linearly independent minimal left ideal of isospin-down elementary fermions $S^d\equiv \mathbb{C}\ell(6)\omega^{\dagger}\omega$. The representations of the minimal ideals are invariant under the electrocolor symmetry $SU(3)_C\times U(1)_{EM}$, whose generators are constructed from the bivectors of the algebra, with each basis state in the ideals transforming as a specific lepton or quark as indicated by their suggestively labeled complex coefficients. That is, the unitary spin transformations that preserve the Witt decomposition are given by $SU(3)_C\times U(1)_{EM}$. 

In terms of the Witt basis ladder operators, the $SU(3)_C$ generators take the form
\begin{eqnarray}
\nonumber\Lambda_1&=&-\alpha_2^{\dagger}\alpha_1-\alpha_1^{\dagger}\alpha_2,\qquad \Lambda_2=i\alpha_2^{\dagger}\alpha_1-i\alpha_1^{\dagger}\alpha_2,\\
\nonumber\Lambda_3&=&\alpha_2^{\dagger}\alpha_2-\alpha_1^{\dagger}\alpha_1,\qquad \Lambda_4=-\alpha_1^{\dagger}\alpha_3-\alpha_3^{\dagger}\alpha_1,\\
\Lambda_5&=&-i\alpha_1^{\dagger}\alpha_3+i\alpha_3^{\dagger}\alpha_1,\qquad \Lambda_6=-\alpha_3^{\dagger}\alpha_2-\alpha_2^{\dagger}\alpha_3,\\
\nonumber\Lambda_7&=&i\alpha_3^{\dagger}\alpha_2-i\alpha_2^{\dagger}\alpha_3,\qquad \Lambda_8=\frac{-1}{\sqrt{3}}(\alpha_1^{\dagger}\alpha_1+\alpha_2^{\dagger}\alpha_2-2\alpha_3^{\dagger}\alpha_3).
\end{eqnarray}
The $U(1)_{EM}$ generator, proportional to the number operator, can be expressed in terms of the Witt basis ladder operators as
\begin{eqnarray}
Q=\frac{1}{3}(\alpha_1^{\dagger}\alpha_1+\alpha_2^{\dagger}\alpha_2+\alpha_3^{\dagger}\alpha_3),
\end{eqnarray}
and gives the electric charge of fermions. 

As an illustrative example we consider $[\Lambda_1, u^g]$:
\begin{eqnarray}
\nonumber\left[ \Lambda_1, u^g\right] &=&\left(-\alpha_2^{\dagger}\alpha_1-\alpha_1^{\dagger}\alpha_2\right) \alpha_1^{\dagger}\alpha_3^{\dagger}-\alpha_1^{\dagger}\alpha_3^{\dagger}\left(-\alpha_2^{\dagger}\alpha_1-\alpha_1^{\dagger}\alpha_2\right),\\
\nonumber &=& -\alpha_2^{\dagger}\alpha_3^{\dagger}\alpha_1\alpha_1^{\dagger}+\alpha_3^{\dagger}\alpha_2^{\dagger}\alpha_1^{\dagger}\alpha_1,\\
\nonumber &=&\alpha_3^{\dagger}\alpha_2^{\dagger}\left( \alpha_1\alpha_1^{\dagger}+\alpha_1^{\dagger}\alpha_1\right) ,\\
&=&\alpha_3^{\dagger}\alpha_2^{\dagger}=u^r.
\end{eqnarray}

%%%%%%%%%%%%%%%%%%%%%%%%%%%%%%%%%%%%%%%%%%
\section{Weak symmetries for one generation of fermions from $\mathbb{C}\ell(4)$}
\subsection{Towards the weak force}

So far we have considered two of the full set of eight minimal left ideals of $\mathbb{C}\ell(6)$. One ideal, $S^u$, consists of isospin-up states, and the other, $S^d$ consists of isospin-down states. The unitary symmetries $SU(3)_C$ and $U(1)_{EM}$ facilitate transitions between states within an ideal. One also notices that multiplying the ideal $S^u$ on the right by $\omega$ changes the ideal into $S^d$. The same is true for multiplying $S^d$ on the right by $\omega^{\dagger}$. Notice that this only works via right-multiplications, not via left-multiplication. Together, $\omega$ and $\omega^{\dagger}$ generate a copy of $\mathbb{C}\ell(2)$, which is a subalgebra of $\mathbb{C}\ell(6)$. The $SU(2)_{\omega}$ generators are written as follows
\begin{eqnarray}\label{w generators}
\tau_1\equiv \omega+\omega^{\dagger},\qquad \tau_2\equiv i(\omega-\omega^{\dagger}),\qquad \tau_3\equiv \omega\omega^{\dagger}-\omega^{\dagger}\omega,
\end{eqnarray}
where it needs to be remembered that this $SU(2)_{\omega}$ acts from the right so that $[\frac{\tau_j}{2}, \frac{\tau_i}{2}]=i\epsilon_{ijk}\frac{\tau_k}{2}$. This $SU(2)_{\omega}$ symmetry effects transitions between states of different ideals. In particular, note that the action of this $SU(2)_{\omega}$ onto a state changes its electric charge by plus or minus one. 

In itself, $\mathbb{C}\ell(2)\cong SU(2)$ generated by $\omega$ and $\omega^{\dagger}$ is not sufficient to describe weak interactions. This is because $SU(2)_{\omega}$ does not correspond to a unitary symmetry that preserves the Witt decomposition of $\mathbb{C}\ell(6)$ into minimal ideals. Nonetheless, this $\mathbb{C}\ell(2)$ plays an important role in weak interactions since a weak doublet contains two particles that belong to different $\mathbb{C}\ell(6)$ minimal left ideals and differ in their electric charge by one. Therefore, to describe the $SU(2)_L$ weak symmetry, we want to include an additional Clifford algebra, whose Witt decomposition is preserved by $SU(2)$. This algebra is $\mathbb{C}\ell(4)$. 

%%%%%%%%%%%%%%%%%%%%%%%%%%%%%%%%%%%%%%%%%%%%%%
%\subsection{Ladder symmetries of $\mathbb{C}\ell(4)$}
\subsection{Minimal right ideals of $\mathbb{C}\ell(4)$ and unitary symmetries}

As for $\mathbb{C}\ell(6)$, the algebra $\mathbb{C}\ell(4)$, with say orthonormal basis $\{w_1, w_2, w_3, w_4\}$, may be rewritten in terms of a nilpotent basis $\{\beta_1,\beta_2,\beta_1^{\dagger},\beta_2^{\dagger}\}$ of ladder operators. The unitary symmetries of these ladder operators turn out to be $SU(2)\times U(1)/ \mathbb{Z}_2\cong U(2)$ \cite{furey2018demonstration}. 

Following the construction of minimal left ideals of $\mathbb{C}\ell(6)$, we here construct the minimal right ideals of $\mathbb{C}\ell(4)$. First define the nilpotents $\Omega=\beta_2\beta_1$ and $\Omega^{\dagger}=\beta_1^{\dagger}\beta_2^{\dagger}$, from which one constructs the idempotents $\Omega\Omega^{\dagger}$ and $\Omega^{\dagger}\Omega$. Two minimal right ideals are then given by $\Omega\Omega^{\dagger}\mathbb{C}\ell(4)$ and $\Omega^{\dagger}\Omega\mathbb{C}\ell(4)$, which are four complex-dimensions. Explicitly the ideals are spanned by the states
\begin{eqnarray}
\{\Omega\Omega^{\dagger},\Omega\Omega^{\dagger}\beta_1,\Omega\Omega^{\dagger}\beta_2,\Omega\Omega^{\dagger}\beta_2\beta_1\},\\
\{\Omega^{\dagger}\Omega,\Omega^{\dagger}\Omega\beta_1^{\dagger},\Omega^{\dagger}\Omega\beta_2^{\dagger},\Omega^{\dagger}\Omega\beta_1^{\dagger}\beta_2^{\dagger}\}.
\end{eqnarray}

The $SU(2)$ and $U(1)$ unitary symmetries that preserve the ladder operator basis of $\mathbb{C}\ell(4)$ are generated by
\begin{eqnarray}\label{SU(2) generators}
T_1&=&-\beta_1\beta_2^{\dagger}-\beta_2\beta_1^{\dagger},\qquad T_2=i\beta_1\beta_2^{\dagger}-i\beta_2\beta_1^{\dagger},\qquad T_3=\beta_1\beta_1^{\dagger}-\beta_2\beta_2^{\dagger},
\end{eqnarray}
and
\begin{eqnarray}
N&=&\beta_1\beta_1^{\dagger}+\beta_2\beta_2^{\dagger}.
\end{eqnarray}
%where again the $SU(2)$ and $U(1)$ symmetries are acting from the right.

%%%%%%%%%%%%%%%%%%%%%%%%%%%%%%%%%%%%%%%%%%%%%%%%%%%%%%%%%%%%%%%%%%%%%%%%%%%%%%%%%%%%%%%%%%%%%%%%%55
%These are the right action generators
%\begin{eqnarray}
%T_1=-\beta_2^{\dagger}\beta_1-\beta_1^{\dagger}\beta_2,\qquad T_2=i\beta_2^{\dagger}\beta_1-i\beta_1^{\dagger}\beta_2,\qquad T_3=\beta_2^{\dagger}\beta_2-\beta_1^{\dagger}\beta_1.
%\end{eqnarray}

Under the action of the $SU(2)$ generators the states $\Omega^{\dagger}\Omega$ and $\Omega^{\dagger}\Omega\beta_1^{\dagger}\beta_2^{\dagger}$ transform as singlets whereas the states $\Omega^{\dagger}\Omega\beta_1^{\dagger}$ and $\Omega^{\dagger}\Omega\beta_2^{\dagger}$ transform into each other as a doublet. The singlet states may then be identified with right handed fermions, and the doublet states with left-handed fermions. Similarly, the states $\Omega\Omega^{\dagger}$ and $\Omega\Omega^{\dagger}\beta_1\beta_2$ are identified with left-handed anti-fermions, whereas $\Omega\Omega^{\dagger}\beta_1$ and $\Omega\Omega^{\dagger}\beta_2$ with right-handed anti-fermions.

Subsequently, we can write two $\mathbb{C}\ell(4)$ minimal right ideals, one for leptons and one for anti-leptons, as
\begin{eqnarray}
L=\nu_R\Omega^{\dagger}\Omega+\nu_L\Omega^{\dagger}\Omega\beta_1^{\dagger}+e^-_L\Omega^{\dagger}\Omega\beta_2^{\dagger}+e^-_R\Omega^{\dagger}\Omega\beta_1^{\dagger}\beta_2^{\dagger},
\end{eqnarray}
and
\begin{eqnarray}
\bar{L}=\bar{\nu}_L\Omega\Omega^{\dagger}+\bar{\nu}_R\Omega\Omega^{\dagger}\beta_1+e^+_R\Omega\Omega^{\dagger}\beta_2+e^+_L\Omega\Omega^{\dagger}\beta_2\beta_1,
\end{eqnarray}
where in both cases the coefficients are complex numbers indicating the particle like which the basis states transforms.

As an example, consider $\left[ T_1, \nu_L\right]: $
\begin{eqnarray}
\nonumber\left[ T_1, \nu_L\right] &=&(-\beta_1\beta_2^{\dagger}-\beta_2\beta_1^{\dagger})\Omega^{\dagger}\Omega\beta_1^{\dagger}-\Omega^{\dagger}\Omega\beta_1^{\dagger}(-\beta_1\beta_2^{\dagger}-\beta_2\beta_1^{\dagger}),\\
\nonumber &=& -\Omega^{\dagger}\Omega\beta_1^{\dagger}(-\beta_1\beta_2^{\dagger}),\\
&=& \Omega^{\dagger}\Omega\beta_2^{\dagger}=e_L.
\end{eqnarray}

Similarly, for the quarks and anti-quarks, the $\mathbb{C}\ell(4)$ minimal right ideals can be written as
\begin{eqnarray}
Q^{(3)}=u^{(3)}_R\Omega^{\dagger}\Omega+u^{(3)}_L\Omega^{\dagger}\Omega\beta_1^{\dagger}+d^{(3)}_L\Omega^{\dagger}\Omega\beta_2^{\dagger}+d^{(3)}_R\Omega^{\dagger}\Omega\beta_1^{\dagger}\beta_2^{\dagger},
\end{eqnarray} 
and
\begin{eqnarray}
\bar{Q}^{(3)}=\bar{u}^{(3)}_L\Omega\Omega^{\dagger}+\bar{u}^{(3)}_R\Omega\Omega^{\dagger}\beta_1+\bar{d}^{(3)}_R\Omega\Omega^{\dagger}\beta_2+\bar{d}^{(3)}_L\Omega\Omega^{\dagger}\beta_2\beta_1,
\end{eqnarray}
where the superscripts refers to the color of the quarks. Presently, as $\mathbb{C}\ell(4)$ minimal right ideals, the quark ideals and lepton ideals are identical. We will be able to distinguish between leptons and quarks once we include the $\mathbb{C}\ell(6)$ minimal left ideals.

\section{Combining $\mathbb{C}\ell(6)$ electrocolor and $\mathbb{C}\ell(4)$ weak states}

We now combine the earlier results based on $\mathbb{C}\ell(6)$ electrocolor states with the $\mathbb{C}\ell(4)$ weak states of the preceding section. We assume for simplicity here that the two algebras commute so that $\alpha_i\beta_j=\beta_j\alpha_i$. Everything that follows still works, with some minor modifications, when $\alpha_i$ and $\beta_j$ anticommute.

The neutrino $\nu$ is represented by the $\mathbb{C}\ell(6)$ minimal left ideal basis state $\omega\omega^{\dagger}$. Via the $\mathbb{C}\ell(4)$ right ideals of the previous section, we can now include chirality. We then have
\begin{eqnarray}
\nu_R&=&\omega\omega^{\dagger}\Omega^{\dagger}\Omega,\\
\nu_L&=&\omega\omega^{\dagger}\Omega^{\dagger}\Omega\beta_1^{\dagger}.
\end{eqnarray}
Similarly, the neutrino's weak doublet partner, the electron $e^-$ in its left- and right-handed form can now be written as
\begin{eqnarray}
e^-_L&=&\alpha_1\alpha_2\alpha_3\omega^{\dagger}\omega\Omega^{\dagger}\Omega\beta_2^{\dagger},\\
e^-_R&=&\alpha_1\alpha_2\alpha_3\omega^{\dagger}\omega\Omega^{\dagger}\Omega\beta_1^{\dagger}\beta_2^{\dagger}.
\end{eqnarray}
Notice that the neutrino and electron live in different $\mathbb{C}\ell(6)$ minimal left ideals, but in the same $\mathbb{C}\ell(4)$ minimal right ideal.

The red up quark $u^r$ with electrocolor symmetry was previously identified with the $\mathbb{C}\ell(6)$ state $\alpha_3^{\dagger}\alpha_2^{\dagger}\omega\omega^{\dagger}$. Via the $\mathbb{C}\ell(4)$ right ideals of the previous section, we can now include chirality. We then have
\begin{eqnarray}
u^r_R&=&\alpha_3^{\dagger}\alpha_2^{\dagger}\omega\omega^{\dagger}\Omega^{\dagger}\Omega,\\
u^r_L&=&\alpha_3^{\dagger}\alpha_2^{\dagger}\omega\omega^{\dagger}\Omega^{\dagger}\Omega\beta_1^{\dagger},
\end{eqnarray}
Subsequently, the red down quark $d^r$ becomes
\begin{eqnarray}
d^r_L&=&\alpha_1\omega^{\dagger}\omega\Omega^{\dagger}\Omega\beta_2^{\dagger},\\
d^r_R&=&\alpha_1\omega^{\dagger}\omega\Omega^{\dagger}\Omega\beta_1^{\dagger}\beta_2^{\dagger}.
\end{eqnarray} 
Again, the red down quark and red up quark belong to different $\mathbb{C}\ell(6)$ ideals but the same $\mathbb{C}\ell(4)$ ideal. %Similarly, the neutrino and up quark (as well as the electron and down quark) belong to the same $\mathbb{C}\ell(6)$ ideal but to different $\mathbb{C}\ell(4)$ ideals.
The antiparticles live in the conjugate $\mathbb{C}\ell(4)$ minimal right ideal. Hence, for example
\begin{eqnarray}
e^+_R&=&\alpha_3^{\dagger}\alpha_2^{\dagger}\alpha_1^{\dagger}\omega\omega^{\dagger}\Omega\Omega^{\dagger}\beta_2,\\
\bar{u}^r_L&=& \alpha_2\alpha_3\omega^{\dagger}\omega\Omega\Omega^{\dagger}.
\end{eqnarray}

In summary, the eight weak-doublets are identified as
\begin{eqnarray}
\begin{pmatrix} \nu_L \\ e^-_L  \end{pmatrix}=\begin{pmatrix} \omega\omega^{\dagger}\Omega^{\dagger}\Omega\beta_1^{\dagger} \\ 
\alpha_1\alpha_2\alpha_3\omega^{\dagger}\omega\Omega^{\dagger}\Omega\beta_2^{\dagger}\end{pmatrix},\qquad
\begin{pmatrix} u^{(3)}_L \\ d^{(3)}_L  \end{pmatrix}=\begin{pmatrix} \alpha_j^{\dagger}\alpha_i^{\dagger}\omega\omega^{\dagger}\Omega^{\dagger}\Omega\beta_1^{\dagger} \\ 
\epsilon_{ijk}\alpha_k\omega^{\dagger}\omega\Omega^{\dagger}\Omega\beta_2^{\dagger}\end{pmatrix},\\
\begin{pmatrix} e^+_R\\ \bar{\nu}_R \end{pmatrix}=\begin{pmatrix} \alpha_3^{\dagger}\alpha_2^{\dagger}\alpha_1^{\dagger}\omega\omega^{\dagger}\Omega\Omega^{\dagger}\beta_2 \\
\omega^{\dagger}\omega\Omega\Omega^{\dagger}\beta_1 \end{pmatrix},\qquad
\begin{pmatrix} \bar{d}^{(3)}_R\\ \bar{u}^{(3)}_R \end{pmatrix}=\begin{pmatrix} \alpha_i^{\dagger}\omega\omega^{\dagger}\Omega\Omega^{\dagger}\beta_2 \\
\epsilon_{ijk}\alpha_j\alpha_k\omega^{\dagger}\omega\Omega\Omega^{\dagger}\beta_1 \end{pmatrix}
\end{eqnarray}
All of the other physical states are weak singlets
\begin{eqnarray}
\left( \nu_R\right) &=&\left( \omega\omega^{\dagger}\Omega^{\dagger}\Omega\right),\quad \left( e^-_R\right) =\left( \alpha_1\alpha_2\alpha_3\omega^{\dagger}\omega\Omega^{\dagger}\Omega\beta_1^{\dagger}\beta_2^{\dagger} \right),\\
\left( e^+_L\right) &=&\left( \alpha_3^{\dagger}\alpha_2^{\dagger}\alpha_1^{\dagger}\omega\omega^{\dagger}\Omega\Omega^{\dagger}\beta_2\beta_1\right),\quad \left( \bar{\nu}_L\right) =\left( \omega^{\dagger}\omega\Omega\Omega^{\dagger} \right),\\
\left( u^{(3)}_R\right) &=&\left( \alpha_j^{\dagger}\alpha_i^{\dagger}\omega\omega^{\dagger}\Omega^{\dagger}\Omega\right),\quad \left( d^{(3)}_R\right) =\left( \epsilon_{ijk}\alpha_k\omega^{\dagger}\omega\Omega^{\dagger}\Omega\beta_1^{\dagger}\beta_2^{\dagger} \right),\\
\left( \bar{d}^{(3)}_L\right) &=&\left( \alpha_i^{\dagger}\omega\omega^{\dagger}\Omega\Omega^{\dagger}\beta_2\beta_1\right),\quad \left( \bar{u}^{(3)}_L\right) =\left( \epsilon_{ijk}\alpha_j\alpha_k\omega^{\dagger}\omega\Omega\Omega^{\dagger} \right).
\end{eqnarray}
%%%%%%%%%%%%%%%%%%%%%%%%%%%%%%%%%%%%%%%%%%%%%5
\section{Weak $SU(2)_L$ symmetries}

Now that we can write down chiral fermions in terms of $\mathbb{C}\ell(6)$ and $\mathbb{C}\ell(4)$ minimal ideals, we must next find appropriate $SU(2)$ generators so that the states transform correctly via the weak symmetry $SU(2)_L$.

Consider the weak doublet consisting of a left handed neutrino and left handed electron. In terms of the ideals we have:
\begin{eqnarray}
\begin{pmatrix} \nu_L \\ e^{-}_L  \end{pmatrix}=\begin{pmatrix} \omega\omega^{\dagger}\Omega^{\dagger}\Omega\beta_1^{\dagger} \\ \alpha_1\alpha_2\alpha_3\omega^{\dagger}\omega\Omega^{\dagger}\Omega\beta_2^{\dagger}  \end{pmatrix}
\end{eqnarray}
To transform the neutrino into the electron requires not only that $\beta_1^{\dagger}$ is transformed into $\beta_2^{\dagger}$ via $T_i$ in equation (\ref{SU(2) generators}), but also that the $\mathbb{C}\ell(6)$ ideal, and electric charge are changed. The latter two transformations are mediated by the $SU(2)_{\omega}$ generators (\ref{w generators}). What is required in the present case then is a combination of the generators (\ref{SU(2) generators}) and (\ref{w generators})\footnote{Herein lies the unique approach of this paper. In \cite{furey20183}, the weak interaction are generated simply by (\ref{SU(2) generators}). This leaves the $\mathbb{C}\ell(6)$ unaffected and the result is that the electron is then represented as $\omega\omega^{\dagger}\Omega^{\dagger}\Omega\beta_2$}. After some deliberation, the suitable $SU(2)_L$ generators can be chosen as
\begin{eqnarray}\label{final SU(2)}
T_1'&\equiv& -\beta_1\beta_2^{\dagger}\omega-\beta_2\beta_1^{\dagger}\omega^{\dagger},\\
T_2'&\equiv& i\beta_1\beta_2^{\dagger}\omega-i\beta_2\beta_1^{\dagger}\omega^{\dagger},\\
T_3'&\equiv& \beta_1\beta_2^{\dagger}\beta_2\beta_1^{\dagger}\omega\omega^{\dagger}-\beta_2\beta_1^{\dagger}\beta_1\beta_2^{\dagger}\omega^{\dagger}\omega.
\end{eqnarray}

As a first example, consider the action of $T_1'$ on the left-handed electron $\nu_L$
\begin{eqnarray}
\nonumber \left[ T_1', \nu_L\right] &=& -(\beta_1\beta_2^{\dagger}\omega+\beta_2\beta_1^{\dagger}\omega^{\dagger})(\omega\omega^{\dagger}\Omega^{\dagger}\Omega\beta_1^{\dagger})+(\omega\omega^{\dagger}\Omega^{\dagger}\Omega\beta_1^{\dagger})(\beta_1\beta_2^{\dagger}\omega+\beta_2\beta_1^{\dagger}\omega^{\dagger}),\\
\nonumber &=& 0+\omega\omega^{\dagger}\Omega^{\dagger}\Omega\beta_1^{\dagger}\beta_1\beta_2^{\dagger}\omega,\\
\nonumber &=& \omega\omega^{\dagger}\omega\Omega^{\dagger}\Omega\beta_2^{\dagger},\\
\nonumber &=& \alpha_1\alpha_2\alpha_3 \omega^{\dagger}\omega\Omega^{\dagger}\Omega\beta_2^{\dagger}= e^-_L.
\end{eqnarray}
Furthermore,
\begin{eqnarray}
\nonumber \left[ T_3', \nu_L\right] &=& - (\omega\omega^{\dagger}\Omega^{\dagger}\Omega\beta_1^{\dagger})(\beta_1\beta_2^{\dagger}\beta_2\beta_1^{\dagger}\omega\omega^{\dagger}-\beta_2\beta_1^{\dagger}\beta_1\beta_2^{\dagger}\omega^{\dagger}\omega),\\
\nonumber &=& - (\omega\omega^{\dagger}\Omega^{\dagger}\Omega\beta_1^{\dagger})\beta_1\beta_2^{\dagger}\beta_2\beta_1^{\dagger}\omega\omega^{\dagger},\\
&=& -\omega\omega^{\dagger}\Omega^{\dagger}\Omega\beta_1^{\dagger}=-\nu_L,
\end{eqnarray}
So that $-\frac{1}{2}T_3'$ returns the correct weak isospin of physical states.

As a second example consider an anti-quark weak doublet.
\begin{eqnarray}
\begin{pmatrix}   \bar{d}^r_R \\ \bar{u}^r_R \end{pmatrix}=\begin{pmatrix} \alpha_1^{\dagger}\omega\omega^{\dagger}\Omega\Omega^{\dagger}\beta_2 \\ 
\alpha_3\alpha_2\omega^{\dagger}\omega\Omega\Omega^{\dagger}\beta_1
 \end{pmatrix}
\end{eqnarray}
Then,
\begin{eqnarray}
\nonumber\left[ T_1', \bar{u}^r_R\right] &=&-(\alpha_3\alpha_2\omega^{\dagger}\omega\Omega\Omega^{\dagger}\beta_1)(-\beta_1\beta_2^{\dagger}\omega-\beta_2\beta_1^{\dagger}\omega^{\dagger}),\\
\nonumber&=& (-\alpha_3\alpha_2\omega^{\dagger})\omega\omega^{\dagger}\Omega\Omega^{\dagger}\beta_2,\\
&=& \alpha_1^{\dagger}\omega\omega^{\dagger}\Omega\Omega^{\dagger}\beta_2=\bar{d}^r_R.
\end{eqnarray}
%and
%\begin{eqnarray}
%\nonumber\left[ T_3',\bar{u}^r_R\right] &=&-(\alpha_3\alpha_2\omega^{\dagger}\omega\Omega\Omega^{\dagger}\beta_1)(\beta_1\beta_2^{\dagger}\beta_2\beta_1^{\dagger}\omega\omega^{\dagger}-\beta_2\beta_1^{\dagger}\beta_1\beta_2^{\dagger}\omega^{\dagger}\omega),\\
%\nonumber&=& -(\alpha_3\alpha_2\omega^{\dagger}\omega\Omega\Omega^{\dagger}\beta_1)(-\beta_2\beta_1^{\dagger}\beta_1\beta_2^{\dagger}),\\
%&=& (\alpha_3\alpha_2\omega^{\dagger}\omega\Omega\Omega^{\dagger}\beta_1=\bar{u}^r_R.
%\end{eqnarray}

%%%%%%%%%%%%%%%%%%%%%%%%%%%%%%%%%%%%%%%%%%%
\section{Including spin degrees of freedom}

In the same way that $\omega$ and $\omega^{\dagger}$ via right multiplication transform between the two different $\mathbb{C}\ell(6)$ minimal left ideals $S^u$ and $S^d$ (with $\omega$ and $\omega^{\dagger}$ generating a $\mathbb{C}\ell(2)\cong SU(2)$ subalgebra, so left multiplication by $\Omega$ and $\Omega^{\dagger}$ facilitates transformations between the two $\mathbb{C}\ell(4)$ minimal right ideals $L$ and $\bar{L}$. Subsequently a copy of $SU(2)_{\Omega}$ can be generated by
\begin{eqnarray}\label{omega generators}
t_1\equiv \Omega+\Omega^{\dagger},\qquad t_2\equiv i(\Omega-\Omega^{\dagger}),\qquad t_3\equiv \Omega\Omega^{\dagger}-\Omega^{\dagger}\Omega.
\end{eqnarray}

Multiplying, for example, a right-handed anti-down quark on the left by $\Omega^{\dagger}$ gives
\begin{eqnarray}
\Omega^{\dagger}\bar{d}^{(3)}_R&=&\Omega^{\dagger}\alpha_i^{\dagger}\omega\omega^{\dagger}\Omega\Omega^{\dagger}\beta_2,\\
&=& \alpha_i^{\dagger}\omega\omega^{\dagger}\Omega^{\dagger}\Omega\beta_1^{\dagger}.
\end{eqnarray}
Thus far, $\alpha_i^{\dagger}\omega\omega^{\dagger}\Omega^{\dagger}\Omega\beta_1^{\dagger}$ has not yet been identified with a physical state. Considering the action of $t_3$ on $\bar{d}^{(3)}$ and $\alpha_i^{\dagger}\omega\omega^{\dagger}\Omega^{\dagger}\Omega\beta_1^{\dagger}$ we find that
\begin{eqnarray}
\left[ t_3, \alpha_i^{\dagger}\omega\omega^{\dagger}\Omega\Omega^{\dagger}\beta_2\right] =+\alpha_i^{\dagger}\omega\omega^{\dagger}\Omega\Omega^{\dagger}\beta_2,\quad \left[ t_3, \alpha_i^{\dagger}\omega\omega^{\dagger}\Omega^{\dagger}\Omega\beta_1^{\dagger}\right] =-\alpha_i^{\dagger}\omega\omega^{\dagger}\Omega^{\dagger}\Omega\beta_1^{\dagger}.
\end{eqnarray}
If we then define $\hat{S}_z=\frac{1}{2}t_3$ we can interpret $\alpha_i^{\dagger}\omega\omega^{\dagger}\Omega\Omega^{\dagger}\beta_2$ as a spin-up  anti-down quark and $\alpha_i^{\dagger}\omega\omega^{\dagger}\Omega^{\dagger}\Omega\beta_1^{\dagger}$ as a spin-down anti-down quark
\begin{eqnarray}
\bar{d}_R^{(3)\uparrow}&=&\alpha_i^{\dagger}\omega\omega^{\dagger}\Omega\Omega^{\dagger}\beta_2\\
\bar{d}_R^{(3)\downarrow}&=&\alpha_i^{\dagger}\omega\omega^{\dagger}\Omega^{\dagger}\Omega\beta_1^{\dagger}
\end{eqnarray}
The same can be done for the other physical states.

%%%%%%%%%%%%%%%%%%%%%%%%%%%%%%%%%%%%%%%%%%%%%%%%%%%%%%%%%%55
\section{$\mathbb{C}\ell(10)$ and $SU(5)$ GUT}

Combining the $\mathbb{C}\ell(6)$ minimal left ideals and $\mathbb{C}\ell(4)$ minimal right ideals, it is possible to rewrite physical states as basis states of the minimal left ideals of $\mathbb{C}\ell(10)$, as is done in \cite{furey20183}. Given our different $SU(2)_L$ generators however means that in the present case the physical states will be represented differently. 

For example, consider the left handed electron (ignoring spin which is absent in \cite{furey20183}). In \cite{furey20183} this state is represented as $e^-_L=\beta_2^{\dagger}\omega^{\dagger}\omega\Omega\Omega^{\dagger}$, where the notation has been adopted to be consistent with the present paper. One sees, due to the lack of $\alpha_1\alpha_2\alpha_3$ that the original $\mathbb{C}\ell(6)$ representation of the electron is not preserved in this construction. Consequently, $Q$ no longer gives the electric charge in this $\mathbb{C}\ell(10)$ scheme. On the other hand, our scheme identifies the left-handed electron as $e^-_L=\alpha_1\alpha_2\alpha_3\omega^{\dagger}\omega\Omega^{\dagger}\Omega\beta_2^{\dagger}$, which may be rewritten as a $\mathbb{C}\ell(10)$ element as $e^-_L=\alpha_1\alpha_2\alpha_3\beta_2^{\dagger}\omega^{\dagger}\omega\Omega\Omega^{\dagger}$. This representation preserved both the electrocolor structure $\alpha_1\alpha_2\alpha_3$ from $\mathbb{C}\ell(6)$ and the weak structure $\beta_2^{\dagger}$ from $\mathbb{C}\ell(4)$. Subsequently $Q$ still gives the correct electric charge of the electron.

The unitary symmetry that preserves a Witt decomposition of $\mathbb{C}\ell(10)$ algebra is $SU(5)$, the basis of the Georgi and Glashow GUT. This GUT predicts additional gauge bosons and associated unobserved physical processes, most famously proton decay. However, in the present case, physical states belong simultaneously to a minimal left ideal of $\mathbb{C}\ell(6)$ and a minimal right ideal of $\mathbb{C}\ell(4)$. Although physical states can be rewritten as basis elements of minimal left ideals of $\mathbb{C}\ell(10)$ they do not span these minimal left ideals. Similarly, the $SU(3)\times U(1)$ symmetry that preserves the $\mathbb{C}\ell(6)$ Witt decomposition and $SU(2)\times U(1)$ symmetry that preserves the $\mathbb{C}\ell(4)$ Witt decomposition also preserve a Witt decomposition of $\mathbb{C}\ell(10)$. The converse is however not true as not all $SU(5)$ unitary symmetries that preserve the Witt decomposition of $\mathbb{C}\ell(10)$ individually preserve the Witt decompositions of $\mathbb{C}\ell(6)$ and $\mathbb{C}\ell(4)$. It is those unitary symmetries of $SU(5)$ that correspond to unobserved physical processes, and these are therefore algebraically excluded in our construction. 

%%%%%%%%%%%%%%%%%%%%%%%%%%%%%%%%%%%%%%%%%%%
\section{Discussion}

Two minimal left ideals of $\mathbb{C}\ell(6)$, and two minimal right ideals of $\mathbb{C}\ell(4)$ were shown to transform as a generation of fermions under the unbroken electrocolor group $SU(3)_C\times U(1)_{EM}$, and the weak group $SU(2)_L$ respectively. Combining the $\mathbb{C}\ell(6)$ and $\mathbb{C}\ell(4)$ ideal basis states, the leptons and quarks for a single generation transforming correctly under the SM gauge group can be represented, including the chirality and spin of the particles. These combined $\mathbb{C}\ell(6)$ and $\mathbb{C}\ell(4)$ ideal basis states can be embedded into a minimal left ideal of $\mathbb{C}\ell(10)$ in such a way that preserves individually the $\mathbb{C}\ell(6)$ structure and $\mathbb{C}\ell(4)$ structure of physical states. This is an improvement over an earlier model \cite{furey20183} where the individual $\mathbb{C}\ell(6)$ structure and $\mathbb{C}\ell(4)$ structure of physical states is not preserved. This improvement is achieved through redefining the $SU(2)_L$ to include factors of $\omega$ and $\omega^{\dagger}$ in appropriate places to ensure that a transformation within a $\mathbb{C}\ell(4)$ minimal right ideal simultaneously maps between different $\mathbb{C}\ell(6)$ minimal left ideals. 

It is important to notice the conceptual difference between the $SU(2)$ generated from (\ref{w generators}) and the $SU(2)$ generated from (\ref{SU(2) generators}). In the former case, $\omega$ and $\omega^{\dagger}$ span a $\mathbb{C}\ell(2)$ subalgebra of $\mathbb{C}\ell(6)$. This algebra is isomorphic to $SU(2)$ and the generators mediate transitions between the $\mathbb{C}\ell(6)$ ideals $S^u$ and $S^d$. This $SU(2)$ symmetry is not a gauge symmetry as it does not transform between states within the same ideal, but rather between states of different ideals. Consequently there are no gauge bosons associated with this $SU(2)$ spin symmetry. In the latter case, the $SU(2)$ generators are constructed from the bivectors of $\mathbb{C}\ell(4)$. This copy of $SU(2)$ contains the symmetries that preserves the Witt decomposition of $\mathbb{C}\ell(4)$. 

In our model, unitary symmetries generated from combinations of $\mathbb{C}\ell(6)$ and $\mathbb{C}\ell(4)$ ladder operators, such as $\alpha_i\beta_j^{\dagger}+\beta_j\alpha^{\dagger}$, are naturally excluded because they do not individually preserve the Witt decompositions of $\mathbb{C}\ell(6)$ and $\mathbb{C}\ell(4)$ ideals\footnote{This excludes the factors of $\omega$ and $\omega^{\dagger}$ in (\ref{final SU(2)}).}. It is precisely these generators that are responsible for the unobserved particle processes, including proton decay, in the Georgi-Glashow $SU(5)$ GUT. The present model therefore naturally excludes all of these unobserved processes.

The motivation for this work came from the results of \cite{gresnigt2018braids} which showed that there is a one-to-one correspondence between the basis states of the minimal left ideals of $\mathbb{C}\ell(6)$ and the braided states in the topological model proposed in \cite{Bilson-Thompson2005}\footnote{For additional work related to this braid model and its connections with division algebras and Clifford algebras, the reader is referred to \cite{Bilson-Thompson2007a,Bilson-Thompson2009,gresnigt2019braided,gresnigt2017braidsgroups,asselmeyer2019braids,cartin2015braids}.}. To be able to extend this curious structural similarity to include the weak force, the underlying electrocolor structure generated from $\mathbb{C}\ell(6)$ should remain intact. This is not the case in the $\mathbb{C}\ell(10)$ representation in \cite{furey20183}, but is the case for the construction considered here. The extension of \cite{gresnigt2018braids} to include the weak force will be the focus of an upcoming paper.

 %%%%%%%%%%%%%%%%%%%%%%%%%%%%%%%%%%%%%%%%%%%%%%%%
\section*{Acknowledgments}

This work is supported by the Natural Science Foundation of the Jiangsu Higher Education Institutions of China Programme grant 19KJB140018 and XJTLU REF-18-02-03 grant.

%\begin{thebibliography}{99}
%\section*{References}
%%%%%%%%%%%%%%%%%%%%%%%%%%%%%%%%%%%%%%%%%%%%%%%%%%%%%%%%%%%%%%%%%%%%%
\bibliography{NielsReferences}  % Replace xxx by your  usercode (no extension)
\bibliographystyle{unsrt}  
%%%%%%%%%%%%%%%%%%%%%%%%%%%%%%%%%%%%%%%%%%%%%%%%%%%%%%%%%%%%%%%%%%%%%

\end{document}